\documentclass[12pt]{article}
\usepackage{graphicx}

\begin{document}

\title{Equivalence of a compressible inviscid flow and the Bloch vector under the thermal Jaynes-Cummings model}

\author{Hiroo Azuma${}^{1,}$\thanks{Email: hiroo.azuma@m3.dion.ne.jp}
\ \ 
and
\ \ 
Masashi Ban${}^{2,}$\thanks{Email: m.ban@phys.ocha.ac.jp}
\\
\\
{\small ${}^{1}$Advanced Algorithm \& Systems Co., Ltd.,}\\
{\small 7F Ebisu-IS Building, 1-13-6 Ebisu, Shibuya-ku, Tokyo 150-0013, Japan}\\
{\small ${}^{2}$Graduate School of Humanities and Sciences, Ochanomizu University,}\\
{\small 2-1-1 Ohtsuka, Bunkyo-ku, Tokyo 112-8610, Japan}
}

\date{\today}

\maketitle

\begin{abstract}
In this paper,
we show that the time evolution of the Bloch vector governed by the thermal Jaynes-Cummings model
is equivalent to a compressible inviscid flow with zero vorticity.
Because of its quasiperiodicity,
the dynamics of the Bloch vector includes countably infinite angular momenta as integrals of motion.
Moreover, to derive the Bloch vector,
we trace out the Hilbert space of the cavity field and remove entanglement between the single atom and the cavity mode.
These facts indicate that the dynamics of the Bloch vector can be described
with a hidden-variable model
that has local determinism and a countably infinite number of degrees of freedom.
Our results fit these considerations.
\end{abstract}

\section{\label{section-introduction}Introduction}
In 1963,
the Jaynes-Cummings model (JCM) was proposed for describing interaction
between a single two-state atom and a single near-resonant quantized cavity mode
\cite{Jaynes1963,Shore1993,Louisell1973,Barnett1997,Schleich2001}.
Considering the Hamiltonian for a magnetic dipole in a magnetic field,
assuming near resonance and applying the rotating-wave approximation to it,
we obtain the Hamiltonian of the JCM.
The JCM has been attracting a lot of researchers' attention in the field of the quantum optics
since 1960s,
because it is soluble and fully quantum mechanical.

One of the most remarkable properties of the JCM is
a phenomenon of collapse and revival of Rabi oscillations.
If we put the cavity field into the coherent state initially,
the JCM shows the spontaneous periodic collapse and revival of the Rabi oscillations in the atomic population inversion
during its time evolution.
This phenomenon is regarded as an evidence for discreteness of photons
and thus for the quantum nature of the radiation.

However, if we put the two-level atom into a certain pure state
and the cavity field into a mixed state in thermal equilibrium at initial time,
we hardly predict how the system evolves under the Jaynes-Cummings (JC) interaction
\cite{Knight1986,Chumakov1993,Azuma2008}.
The thermal fluctuation of the radiation field lets the Bloch vector develop in time
in a confusing manner.
Both its norm and direction change hard at random,
so that it seems to be in disorder.
When we discuss the thermal JCM,
we have to handle an intractable infinite series.
If the single cavity field is resonant with the atom,
the $n$th term of this intractable infinite series is given by a trigonometric function of
$\sqrt{n}t$ for $n=0, 1, 2, ...$,
where $t$ represents the variable of time.
Because it cannot be a Fourier series,
the sum of the series varies in an unpredictable manner
as the time $t$ progresses.

Hioe {\it et al}. investigate the long-term behaviour of the JCM from a statistical point of view
\cite{Hioe1983}.
They find that there exists a mean angular frequency characterizing
the highly irregular and random oscillations of the atomic variables in both cases
when the initial photon distributions of the cavity field are given by a coherent state and a thermal equilibrium state.

Yoo {\it et al}. investigate the long-term behaviour of collapse and revivals of the Rabi oscillations in the JCM \cite{Yoo1981}.
The revivals become less complete and begin to broaden,
eventually overlapping each other at still longer times.
The characteristic timescale of this observation is on the order of
$6\pi\leq \tau\leq 20\pi$,
where $\tau=gt/(2\sqrt{\bar{n}})$,
$g$ is a coupling constant of the JC interaction,
$t$ represents the variable of time and $\bar{n}$ is an average photon number.
The saddle point approach is performed to the atomic inversion and interference between revival signals is examined.
Their systematic analysis is related to the notion of fractional revivals,
which is pointed out in Ref.~\cite{Averbukh1989}.

Averbukh and Perelman study the long-term evolution of quantum wave packets and establish the concept of the fractional revivals
\cite{Averbukh1989}.
They indicate that the fractional revivals can be found in a detailed numerical study of the Rydberg wave packets \cite{Parker1986}.
The fractional revivals are also observed in the time evolution of the atomic inversion for the JCM
with initial coherent state of the cavity field \cite{Schleich2001}.

A.A.~Karatsuba and E.A.~Karatsuba study the JC sum,
which determines the atomic inversion \cite{Karatsuba2007,Karatsuba2009a,Karatsuba2009b,Karatsuba2009c}.
In Ref.~\cite{Karatsuba2007},
an approximation of  the JC sum is evaluated with the theorem on the approximation of a trigonometric sum by a shorter one.
Dealing with the problem in this way,
main contributions are extracted as a finite series from the JC sum.
In Refs.~\cite{Karatsuba2009a,Karatsuba2009b,Karatsuba2009c},
the JC sum for a coherent state of a single cavity mode is investigated.
Asymptotic formulae of it are derived using the functional equation for the Jacobi Theta-functions.
The obtained formulae make it be possible to predetermine the details of the behaviour of the inversion
and to describe JC collapses and revivals on various time intervals.

In Ref.~\cite{Azuma2014},
Azuma and Ban show that the time evolution of the Bloch vector under the thermal JCM
has a quasiperiodic structure.
Because of the quasiperiodicity,
the dynamics of the Bloch vector has countably infinite incommensurate angular frequencies,
$\omega_{n}=\sqrt{n}$ for $n=1, 2, 3, 5, ...$,
where the variable $n$ comes from the number of photons of a cavity mode
and the series $n=1, 2, 3, 5, ...$ lets $\{\omega_{n}\}$ be incommensurate.
This implies that the system has countably infinite angular momenta
as integrals of motion,
$I_{n}$ for $n=1, 2, 3, 5, ...$.
Thus, we have to conclude that we cannot describe the motion of the Bloch vector
as a system with a finite number of degrees of freedom,
for example,
the $n$-body problem in classical mechanics.
Hence, we can expect the motion of the Bloch vector
to be equivalent to a hidden-variable model that has a countably infinite number of degrees of freedom \cite{Bell1987}.
In this paper,
we regard a hidden-variable model as a model of classical theory in contrast to quantum theory.

In the above paragraph,
we suppose that the dynamics of the Bloch vector has to be written as a hidden-variable model.
The reason why is as follows.
The dynamics of the Bloch vector is derived from the Schr\"{o}dinger equation
with the JC Hamiltonian.
For describing an explicit form of the Bloch vector,
what we apply to the quantum mechanical state in fact
are only providing the initial state of the cavity field
as a density matrix of a mixed state and tracing out its Hilbert space.
Because we trace out the Hilbert space of the cavity field,
the entanglement between the single atom and the cavity field is removed.
Thus, the dynamics of the Bloch vector has to be described with a hidden-variable model that exhibits local determinism.
Maxwell's electromagnetic theory and the fluid dynamics give us typical examples of hidden-variable theory \cite{Landau1975,Landau1987}.

Here, we examine the meanings of the hidden-variable model precisely.
The initial state of the atom and the cavity photons is given by
\begin{equation}
\rho_{\mbox{\scriptsize AP}}(0)=\rho_{\mbox{\scriptsize A}}(0)\otimes\rho_{\mbox{\scriptsize P}}(0),
\label{separable-state-0}
\end{equation}
where $\rho_{\mbox{\scriptsize A}}(0)$ and $\rho_{\mbox{\scriptsize P}}(0)$ are density operators of the atom and the cavity photons,
respectively.
[The indices A and P stand for the atom and the photon, respectively.]
Then, the time evolution under the JC Hamiltonian lets the subsystems of the atom and the cavity photons be entangled.
This implies that the whole density operator is not generally given by
\begin{equation}
\rho_{\mbox{\scriptsize AP}}(t)
=
\sum_{i}w_{i}\rho_{\mbox{\scriptsize A},i}(t)\otimes\rho_{\mbox{\scriptsize P},i}(t),
\label{separable-state-1}
\end{equation}
where $w_{i}\geq 0$ $\forall i$ and $\sum_{i}w_{i}=1$.
The density operator of Eq.~(\ref{separable-state-1}) is called classically correlated
and there are no quantum correlations between the atom and the photons.
It is indicated that the system is described with quantum logic in most cases
if its density operator is not given by the form of Eq.~(\ref{separable-state-1}) \cite{Peres1995,Werner1989}.
To obtain the Bloch vector,
we trace out the Hilbert space of the cavity photons as
\begin{equation}
\rho'_{\mbox{\scriptsize A}}(t)
=
\mbox{Tr}_{\mbox{\scriptsize P}}\rho_{\mbox{\scriptsize AP}}(t),
\end{equation}
where $\rho_{\mbox{\scriptsize AP}}(t)$ is not given by the form of Eq.~(\ref{separable-state-1}).
Because quantum entanglement between the atom and the cavity photons  is removed,
$\rho'_{\mbox{\scriptsize A}}(t)$ is described with a hidden-variable model \cite{Bell1987}.
From these considerations,
we conclude that the Bloch vector can be described with a classical model.
Moreover, the Bloch vector has a countably infinite number of degrees of freedom.
Hence, the Bloch vector has to be equivalent to a hidden-variable model with a countably infinite number of degrees of freedom,
which admits Maxwell's electromagnetic theory, the fluid dynamics and other classical field theories.

On the one hand,
in Ref.~\cite{Azuma2014},
we consider quasiperiodicity to be the most important factor
in understanding confusion of the Bloch vector's trajectories under the thermal JCM.
On the other hand,
in this paper,
we try to find physical meanings of it by drawing analogy between the Bloch vector and a compressible fluid.
We show that we can rewrite the time evolution of the Bloch vector under the thermal JCM
as a compressible inviscid flow
with zero vorticity.
The current paper is a sequel of Ref.~\cite{Azuma2014}.

This paper is organized as follows.
In Sec.~\ref{section-review-TJCM},
we give a brief review of the thermal JCM.
In Sec.~\ref{section-equivalence-Bloch-vector-compressible-fluid},
we show the equivalence of the Bloch vector and a compressible fluid.
In Sec.~\ref{section-physical-meanings}, we give physical meanings of quantities of the fictitious fluid,
for instance, the density $\rho$,
the pressure $p$ and the external force per unit mass $\mbox{\boldmath $K$}$,
which we introduce in Sec.~\ref{section-equivalence-Bloch-vector-compressible-fluid}.
In Sec.~\ref{section-discussions},
we give brief discussions.
In Appendix~\ref{section-appendix-A},
we consider how to build the Hamiltonian, which yields dynamics of the compressible inviscid fluid
with zero vorticity.
Moreover, we consider the reason why the trajectory of the Bloch vector appears to intersect itself
as shown in Sec.~\ref{section-review-TJCM}.

\section{\label{section-review-TJCM}A brief review of the thermal JCM}
In this section,
we give a brief review of the thermal JCM.
To describe its dynamics,
we use the notation of Refs.~\cite{Azuma2008,Azuma2014}.
The original Hamiltonian of the JCM is expressed in the form,
\begin{equation}
H=\frac{\hbar}{2}\omega_{0}\sigma_{z}+\hbar\omega a^{\dagger}a
+\hbar g(\sigma_{+}a+\sigma_{-}a^{\dagger}),
\label{JCM-Hamiltonian}
\end{equation}
\begin{equation}
\sigma_{\pm}=\frac{1}{2}(\sigma_{x}\pm i\sigma_{y}),
\label{definition-sigma+-}
\end{equation}
\begin{equation}
\sigma_{x}=
\left(
\begin{array}{cc}
0 & 1 \\
1 & 0
\end{array}
\right),
\quad
\sigma_{y}=
\left(
\begin{array}{cc}
0 & -i \\
i & 0
\end{array}
\right),
\quad
\sigma_{z}=
\left(
\begin{array}{cc}
1 & 0 \\
0 & -1
\end{array}
\right),
\label{definition-Pauli-matrices}
\end{equation}
\begin{equation}
[a,a^{\dagger}]=1,
\quad\quad
[a,a]=[a^{\dagger},a^{\dagger}]=0,
\label{commutation-relations-creation-annihilation-operators-photons}
\end{equation}
where the Pauli matrices act on atomic state vectors,
and $a^{\dagger}$ and $a$ are photon creation and annihilation operators, respectively.
Here, we put the whole system into the following initial state for $t=0$:
\begin{equation}
\rho_{\mbox{\scriptsize AP}}(0)
=\rho_{\mbox{\scriptsize A}}(0)\otimes\rho_{\mbox{\scriptsize P}},
\label{whole-system-initial-state}
\end{equation}
\begin{equation}
\rho_{\mbox{\scriptsize A}}(0)
=\sum_{i,j\in\{0,1\}}\rho_{\mbox{\scriptsize A},ij}(0)
|i\rangle_{\mbox{\scriptsize A}}{}_{\mbox{\scriptsize A}}\langle j|,
\label{atom-initial-state}
\end{equation}
\begin{equation}
\rho_{\mbox{\scriptsize P}}
=
(1-e^{-\beta\hbar\omega})\exp(-\beta\hbar\omega a^{\dagger}a),
\label{field-initial-state}
\end{equation}
\begin{equation}
|0\rangle_{\mbox{\scriptsize A}}
=
\left(
\begin{array}{c}
1 \\
0
\end{array}
\right),
\quad
|1\rangle_{\mbox{\scriptsize A}}
=
\left(
\begin{array}{c}
0 \\
1
\end{array}
\right),
\label{atom-basis-vector}
\end{equation}
where $\{\rho_{\mbox{\scriptsize A},ij}\}$ represent
an arbitrary $2\times 2$ Hermitian matrix
with nonnegative eigenvalues and trace unity.
The density operator $\rho_{\mbox{\scriptsize P}}$ given by Eq.~(\ref{field-initial-state})
represents the thermal equilibrium state weighted by the Bose-Einstein distribution
for the inverse of the temperature $\beta[=1/(k_{\mbox{\scriptsize B}}T)]$.

From now on, for simplicity,
we assume $\Delta\omega=\omega-\omega_{0}=0$.
Then, we can write down the Bloch vector
$\mbox{\boldmath $S$}(t)=(S_{x}(t),S_{y}(t),S_{z}(t))$
as follows:
\begin{equation}
\rho_{\mbox{\scriptsize A}}(t)
=
\frac{1}{2}
[\mbox{\boldmath $I$}+\mbox{\boldmath $S$}(t)\cdot \mbox{\boldmath $\sigma$}],
\end{equation}
\begin{equation}
\mbox{\boldmath $S$}(t)
=
\left(
\begin{array}{ccc}
L_{1}(t) & 0 & 0 \\
0 & L_{1}(t) & 0 \\
0 & 0 & L_{3}(t)
\end{array}
\right)
\mbox{\boldmath $S$}(0)
+
\left(
\begin{array}{c}
0\\
0\\
L_{4}(t)
\end{array}
\right),
\label{evolution-Bloch-vector-resonant}
\end{equation}
\begin{eqnarray}
L_{1}(t)&=&
(1-e^{-\beta})
\sum_{n=0}^{\infty}
\cos(\sqrt{n+1}t)\cos(\sqrt{n}t)e^{-n\beta}, \nonumber \\
L_{3}(t)&=&
\frac{1}{2}(1-e^{-\beta})
+\frac{e^{2\beta}-1}{2e^{\beta}}
\sum_{n=1}^{\infty}
\cos(2\sqrt{n}t)e^{-n\beta}, \nonumber \\
L_{4}(t)&=&
-\frac{1}{2}(1-e^{-\beta})
+\frac{(e^{\beta}-1)^{2}}{2e^{\beta}}
\sum_{n=1}^{\infty}
\cos(2\sqrt{n}t)e^{-n\beta},
\label{definition-L1-L3-L4}
\end{eqnarray}
where we assume $\omega\neq 0$ and $g\neq 0$,
and we replace parameters $\beta\hbar\omega$ and $t|g|$
with $\beta$ and $t$, respectively.
They imply that the time $t$ is in units of $|g|^{-1}$ and
the inverse of the temperature $\beta$ is in units of $(\hbar\omega)^{-1}$.
As a result of these replacements,
the Bloch vector $\mbox{\boldmath $S$}(t)$ depends only on two dimensionless variables, $t$ and $\beta$.
Derivation of Eqs.~(\ref{evolution-Bloch-vector-resonant})
and (\ref{definition-L1-L3-L4}) is given in Ref.~\cite{Azuma2014}.

\begin{figure}
\begin{center}
\includegraphics[scale=1.0]{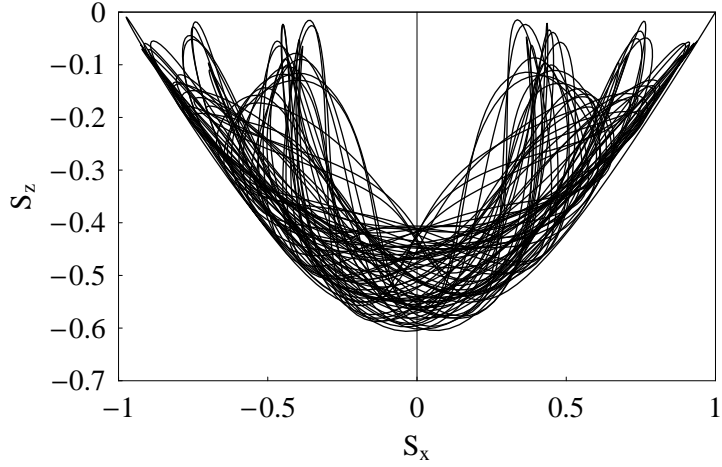}
\end{center}
\caption{A trajectory of $\mbox{\boldmath $S$}(t)$ given
by Eqs.~(\ref{definition-L1-L3-L4}) and (\ref{Bloch-xz-plane})
for $0\leq t \leq 250$ and $\beta=1.0$.
The horizontal and vertical axes represent $S_{x}$ and $S_{z}$, respectively.}
\label{Figure01}
\end{figure}

Here, putting the initial state of the atom into
$(1/\sqrt{2})(|0\rangle_{\mbox{\scriptsize A}}+|1\rangle_{\mbox{\scriptsize A}})$,
we obtain the initial Bloch vector $\mbox{\boldmath $S$}(0)=(1,0,0)$
and its time evolution
\begin{equation}
\mbox{\boldmath $S$}(t)
=
\left(
\begin{array}{c}
L_{1}(t)\\
0\\
L_{4}(t)
\end{array}
\right).
\label{Bloch-xz-plane}
\end{equation}
Equation~(\ref{Bloch-xz-plane}) tells us that the Bloch vector $\mbox{\boldmath $S$}(t)$ always lies
on the $xz$-plane $\forall t(\geq 0)$,
so that it is convenient for tracing the trajectory of $\mbox{\boldmath $S$}(t)$
as time passes.
Thus, in Figs.~\ref{Figure01}, \ref{Figure02}, \ref{Figure03}, \ref{Figure04} and \ref{Figure05},
we only examine the case where the Bloch vector is given
by Eqs.~(\ref{definition-L1-L3-L4}) and (\ref{Bloch-xz-plane}) numerically.

Figure~\ref{Figure01} shows a trajectory of the Bloch vector $\mbox{\boldmath $S$}(t)$
given by Eqs.~(\ref{definition-L1-L3-L4}) and (\ref{Bloch-xz-plane})
for $\beta=1.0$ and $0\leq t\leq 250$.
To obtain Fig.~\ref{Figure01}, we replace an infinite summation $\sum_{n=0}^{\infty}$
with a finite summation $\sum_{n=0}^{150}$
for an actual numerical calculation of $L_{1}(t)$ defined in Eq.~(\ref{definition-L1-L3-L4}).
For calculating $L_{4}(t)$ numerically,
we give a similar treatment.
Throughout this paper,
whenever we carry out numerical calculations of $L_{1}(t)$ and $L_{4}(t)$,
we always apply this approximation to them.

Here, we evaluate numerical errors caused by the above treatment \cite{Azuma2014}.
For example, we can estimate the upper bound of numerical errors for $L_{1}(t)$ as
\begin{eqnarray}
E[L_{1}]
&\leq&
(1-e^{-\beta})
\Biggl|
\sum_{n=151}^{\infty}\cos(\sqrt{n+1}t)\cos(\sqrt{n}t)e^{-n\beta}
\Biggr|
\nonumber \\
&\leq&
(1-e^{-\beta})\sum_{n=151}^{\infty}e^{-n\beta} \nonumber \\
&=&
(1-e^{-\beta})e^{-151\beta}\sum_{n=0}^{\infty}e^{-n\beta} \nonumber \\
&=&
e^{-151\beta}.
\label{L-1-error-estimation}
\end{eqnarray}
Thus, if we assume $0.5\leq\beta\leq 5.0$,
we obtain $E[L_{1}]\leq 1.63\times 10^{-33}$.
Similar things hold for $L_{4}(t)$.
In Refs.~\cite{Karatsuba2007,Karatsuba2009a,Karatsuba2009b,Karatsuba2009c},
the following is pointed out.
In the series included in $L_{1}(t)$ and $L_{4}(t)$,
it is important to distinguish terms which give the major contributions and terms which give the minor ones.
However, such precise analyses are beyond the purpose of this paper and we conclude Eq.~(\ref{L-1-error-estimation}) is enough.

Closing this section,
we note the following two facts.
First, $L_{1}(t)$ has countably infinite incommensurate angular frequencies,
$\omega_{n}=\sqrt{n}$ for $n=1, 2, 3, 5, ...$.
This means that the system has countably infinite angular momenta as integrals of motion,
$I_{n}$ for $n=1, 2, 3, 5, ...$
because of its quasiperiodicity.
[$L_{3}(t)$ and $L_{4}(t)$ have countably infinite incommensurate angular frequencies,
$\omega'_{n}=2\sqrt{n}$ for $n=1, 2, 3, 5, ...$, and we obtain a similar observation.]
Thus, we have to conclude that we cannot describe the motion of the Bloch vector
as the $n$-body problem in classical mechanics,
which has a finite number of degrees of freedom.

Second, we derive the Bloch vector $\mbox{\boldmath $S$}(t)$
given by Eqs.~(\ref{evolution-Bloch-vector-resonant}) and (\ref{definition-L1-L3-L4})
from the
\\
Schr{\"o}dinger equation,
whose Hamiltonian is defined
in Eqs.~(\ref{JCM-Hamiltonian}), (\ref{definition-sigma+-}),
(\ref{definition-Pauli-matrices}) and (\ref{commutation-relations-creation-annihilation-operators-photons}),
originally.
As explained in the former half of this section,
for obtaining $\mbox{\boldmath $S$}(t)$,
what we apply to the quantum mechanical state in fact are only
providing the initial state of the cavity field at $t=0$ as a density matrix of a mixed state and
tracing out its Hilbert space $\forall t>0$.
Because we trace out the Hilbert space of the cavity field,
the entanglement between the single atom and the cavity field is removed.
Thus, the dynamics of the Bloch vector has to be described with a hidden-variable model with local determinism.

Hence,
we can expect the motion of the Bloch vector to be equivalent to a hidden-variable model that has a countably infinite number of degrees of freedom.
In fact, we confirm that the motion of the Bloch vector is equivalent to the dynamics
of a compressible fluid in Sec.~\ref{section-equivalence-Bloch-vector-compressible-fluid}.

\section{\label{section-equivalence-Bloch-vector-compressible-fluid}
Equivalence between the Bloch vector and a compressible fluid}
In this section,
we discuss equivalence between the dynamics of the Bloch vector and a compressible flow.
We demonstrate that the trajectory of the Bloch vector is exactly equivalent to hydrodynamics of a compressible fluid,
which is inviscid and has zero vorticity
\cite{Landau1987,Lamb1932}.

Now, we describe $\mbox{\boldmath $S$}(t)$ as $\mbox{\boldmath $x$}(t)=(x(t),y(t),z(t))$.
Thus, we rewrite Eq.~(\ref{evolution-Bloch-vector-resonant}) as
\begin{equation}
\mbox{\boldmath $x$}(t)
=
\left(
\begin{array}{c}
x(t) \\
y(t) \\
z(t)
\end{array}
\right)
=
\left(
\begin{array}{c}
L_{1}(t)x(0) \\
L_{1}(t)y(0) \\
L_{3}(t)z(0)+L_{4}(t)
\end{array}
\right),
\label{evolution-fluid-field}
\end{equation}
where $|\mbox{\boldmath $x$}(0)|\leq 1$.
We consider Eq.~(\ref{evolution-fluid-field}) to be the Lagrangian forms of the dynamical equations
of a fluid.
In Eq.~(\ref{evolution-fluid-field}),
we let $\mbox{\boldmath $x$}(0)$ be the initial co-ordinates of any particle of the fluid
and $\mbox{\boldmath $x$}(t)$ be its co-ordinates at time $t$.
Hence, Eq.~(\ref{evolution-fluid-field}) expresses the time evolution of the fluid particles
being put initially inside a unit circle,
that is, $|\mbox{\boldmath $x$}(0)|\leq 1$.

From Eq.~(\ref{evolution-fluid-field}),
the velocity of the fluid particle at point $\mbox{\boldmath $x$}(t)$ at time $t$
is given by
\begin{equation}
\dot{\mbox{\boldmath $x$}}(t)
=
\left(
\begin{array}{c}
\dot{x}(t) \\
\dot{y}(t) \\
\dot{z}(t)
\end{array}
\right)
=
\left(
\begin{array}{c}
\dot{L}_{1}(t)x(0) \\
\dot{L}_{1}(t)y(0) \\
\dot{L}_{3}(t)z(0)+\dot{L}_{4}(t)
\end{array}
\right).
\label{evolution-fluid-field-velocity}
\end{equation}
Thus, from Eqs.~(\ref{definition-L1-L3-L4}) and (\ref{evolution-fluid-field-velocity}),
calculating an initial velocity at every point inside the unit circle explicitly,
we obtain
\begin{eqnarray}
\mbox{\boldmath $v$}(t=0;x(0),y(0),z(0))
&=&
\dot{\mbox{\boldmath $x$}}(t=0;x(0),y(0),z(0)) \nonumber \\
&=&
\mbox{\boldmath $0$}
\quad\quad
\mbox{$\forall \mbox{\boldmath $x$}(0)$ such that $|\mbox{\boldmath $x$}(0)|\leq 1$}.
\end{eqnarray}
To obtain the dynamical equations of the fluid,
we rewrite Eq.~(\ref{evolution-fluid-field}) as the following relations,
\begin{equation}
x(0)=\frac{x(t)}{L_{1}(t)},
\quad
y(0)=\frac{y(t)}{L_{1}(t)},
\quad
z(0)=\frac{z(t)-L_{4}(t)}{L_{3}(t)}.
\label{relations-position-initial-time-t}
\end{equation}
Substitution of Eq.~(\ref{relations-position-initial-time-t})
into Eq.~(\ref{evolution-fluid-field-velocity})
yields
\begin{equation}
\dot{\mbox{\boldmath $x$}}(t)
=
\left(
\begin{array}{c}
{[}\dot{L}_{1}(t)/L_{1}(t){]}x(t) \\
{[}\dot{L}_{1}(t)/L_{1}(t){]}y(t) \\
{[}\dot{L}_{3}(t)/L_{3}(t){]}[z(t)-L_{4}(t)]+\dot{L}_{4}(t)
\end{array}
\right).
\label{evolution-fluid-field-velocity-B}
\end{equation}
Therefore, the velocity of the fluid at point $\mbox{\boldmath $x$}=(x,y,z)$ at time $t$ is
described as
\begin{equation}
\mbox{\boldmath $v$}(t,x,y,z)
=
\left(
\begin{array}{c}
{[}\dot{L}_{1}(t)/L_{1}(t){]}x \\
{[}\dot{L}_{1}(t)/L_{1}(t){]}y \\
{[}\dot{L}_{3}(t)/L_{3}(t){]}[z-L_{4}(t)]+\dot{L}_{4}(t)
\end{array}
\right).
\label{evolution-fluid-field-velocity-C}
\end{equation}

Here, let us point out some characteristic properties of the fluid
obtained from the Bloch vector.
From Eq.~(\ref{evolution-fluid-field-velocity-C}), we obtain
\begin{equation}
\nabla\cdot\mbox{\boldmath $v$}
=
2\frac{\dot{L}_{1}(t)}{L_{1}(t)}+\frac{\dot{L}_{3}(t)}{L_{3}(t)}.
\label{nabla-velocity-1}
\end{equation}
Thus, we understand that $\nabla\cdot\mbox{\boldmath $v$}$ depends only on $\beta$ and $t$.
By numerical calculations, we can confirm $\nabla\cdot\mbox{\boldmath $v$}\neq 0$
at almost all times $t$ for $0<\beta<+\infty$.
Thus, we consider that the following relation holds in general,
\begin{equation}
\nabla\cdot\mbox{\boldmath $v$}
\neq
0.
\label{divergence-0}
\end{equation}
In addition, from Eq.~(\ref{evolution-fluid-field-velocity-C}),
we can obtain the following equation, with ease:
\begin{equation}
\nabla\times\mbox{\boldmath $v$}
=
\mbox{\boldmath $0$}
\quad\quad
\forall(t,\mbox{\boldmath $x$}).
\label{rotation-0}
\end{equation}

Then, let us think about the Eulerian equation of continuity,
\begin{equation}
\frac{\partial \rho}{\partial t}
+
\nabla\cdot(\rho\mbox{\boldmath $v$})=0.
\label{equation-continuity}
\end{equation}
If we assume that the fluid is incompressible,
that is, $\rho=\mbox{constant}$,
we obtain $\nabla\cdot\mbox{\boldmath $v$}=0$
$\forall (t,\mbox{\boldmath $x$})$.
However, it contradicts Eq.~(\ref{divergence-0}),
so that we conclude that the fluid has to be compressible.
Moreover, Eq.~(\ref{rotation-0}) implies that there is no vorticity
in any portion of the fluid at any time.
In general, a concept of turbulence closely relates to randomness and complexity of vorticity.
Thus, the fluid never shows turbulence motion.

Let us examine dynamics of the fluid further.
From Eqs.~(\ref{evolution-fluid-field-velocity-C}) and (\ref{rotation-0}),
we obtain the following velocity-potential $\Phi$:
\begin{equation}
\mbox{\boldmath $v$}=\nabla \Phi,
\label{potential-fluid-field-A}
\end{equation}
\begin{equation}
\Phi
=
\frac{\dot{L}_{1}(t)}{L_{1}(t)}(\frac{x^{2}}{2}+\frac{y^{2}}{2})
+
\frac{\dot{L}_{3}(t)}{L_{3}(t)}[\frac{z^{2}}{2}-zL_{4}(t)]+\dot{L}_{4}(t)z.
\label{potential-fluid-field-B}
\end{equation}
Then, from Eq.~(\ref{potential-fluid-field-B}), we obtain the following relation:
\begin{equation}
\triangle\Phi
=
2\frac{\dot{L}_{1}(t)}{L_{1}(t)}
+
\frac{\dot{L}_{3}(t)}{L_{3}(t)}.
\label{potential-fluid-field-C}
\end{equation}
Thus, using Eq.~(\ref{potential-fluid-field-A}) and (\ref{potential-fluid-field-C}),
we arrive at
\begin{equation}
\triangle\mbox{\boldmath $v$}
=
\triangle\nabla\Phi
=
\nabla\triangle\Phi
=
\mbox{\boldmath $0$}.
\label{potential-fluid-field-D}
\end{equation}
Equation~(\ref{potential-fluid-field-D}) implies that
the viscosity of the fluid has no effects on evolution of an actual motion of a fluid element.
The reason why is given in the following.

In general, the equations of motion of a compressible fluid are given by
\begin{equation}
\frac{D\mbox{\boldmath $v$}}{Dt}
=
\frac{\partial\mbox{\boldmath $v$}}{\partial t}
+
(\mbox{\boldmath $v$}\cdot\nabla)\mbox{\boldmath $v$}
=
\mbox{\boldmath $K$}
-
\frac{1}{\rho}\nabla p
+
\frac{\mu}{\rho}
\triangle\mbox{\boldmath $v$},
\label{Navier-Stokes-equations-0}
\end{equation}
which are called the Navier-Stokes equations.
In Eq.~(\ref{Navier-Stokes-equations-0}),
$\rho$, $p$, $\mbox{\boldmath $K$}$ and $\mu$
represent the density of the fluid,
the pressure, the external force per unit mass and the viscosity of the fluid,
respectively.
However, because of Eq.~(\ref{potential-fluid-field-D}),
we can drop the term including $\mu$ from the Navier-Stokes equations.
Thus, we understand that the viscosity $\mu$ has no effects on the motion of the fluid,
in fact.

Putting the above considerations together,
we obtain the following observations.
If we regard the Bloch vector $\mbox{\boldmath $S$}(t)$ as a fluid,
it is compressible and inviscid.
Moreover, it has no vorticity.
By reason of the fact that the fluid is inviscid and has zero vorticity,
its motion seems to be simple for us.
However, this impression is not true for the compressible fluid.
Indeed, sometimes we can hardly predict the motion of the compressible fluid,
even if it does not have the viscosity.

\begin{figure}
\begin{center}
\includegraphics[scale=1.0]{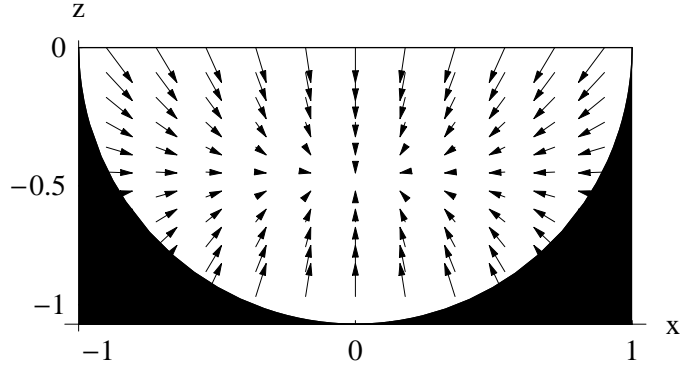}
\end{center}
\caption{This figure shows the velocity $\mbox{\boldmath $v$}(t,\mbox{\boldmath $x$})$
given by Eq.~(\ref{evolution-fluid-field-velocity-C})
at points of $|\mbox{\boldmath $x$}|\leq 1$ and $z\leq 0$ on the $xz$-plane
at time $t=0.5$ with $\beta=1.0$.
The horizontal and vertical axes represent $x$ and $z$, respectively.
We draw the velocity of the fluid $\mbox{\boldmath $v$}(t,\mbox{\boldmath $x$})$
as an arrow (a vector) located at the point $\mbox{\boldmath $x$}$.
The lengths of arrows represent only just the ratios
of their norms compared with each other.}
\label{Figure02}
\end{figure}

\begin{figure}
\begin{center}
\includegraphics[scale=1.0]{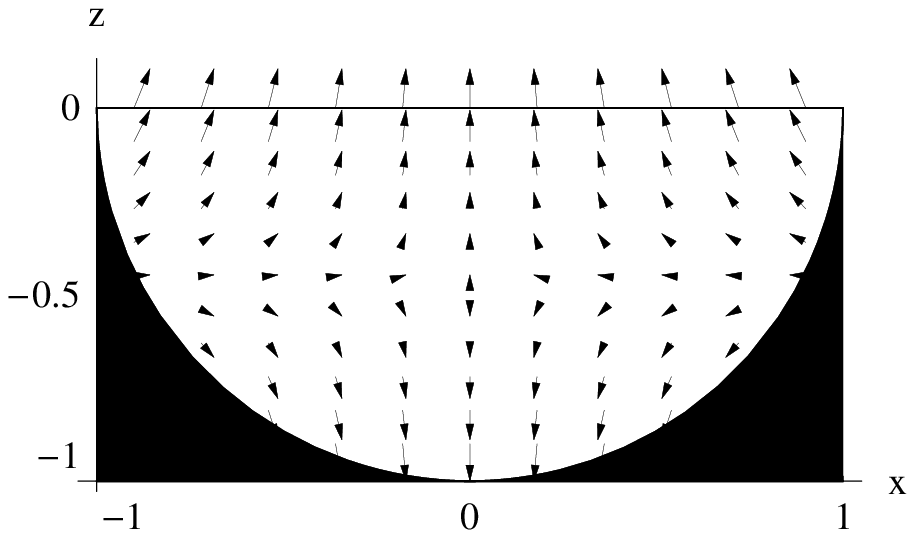}
\end{center}
\caption{This figure shows the velocity $\mbox{\boldmath $v$}(t,\mbox{\boldmath $x$})$
given by Eq.~(\ref{evolution-fluid-field-velocity-C})
at points of $|\mbox{\boldmath $x$}|\leq 1$ and $z\leq 0$ on the $xz$-plane
at time $t=1.0$ with $\beta=1.0$.
The horizontal and vertical axes represent $x$ and $z$, respectively.
We draw the velocity of the fluid $\mbox{\boldmath $v$}(t,\mbox{\boldmath $x$})$
as an arrow (a vector) located at the point $\mbox{\boldmath $x$}$.
The lengths of arrows represent only just the ratios
of their norms compared with each other.}
\label{Figure03}
\end{figure}

\begin{figure}
\begin{center}
\includegraphics[scale=1.0]{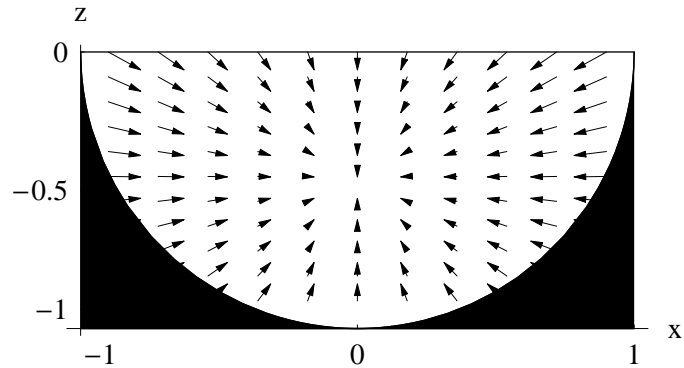}
\end{center}
\caption{This figure shows the velocity $\mbox{\boldmath $v$}(t,\mbox{\boldmath $x$})$
given by Eq.~(\ref{evolution-fluid-field-velocity-C})
at points of $|\mbox{\boldmath $x$}|\leq 1$ and $z\leq 0$ on the $xz$-plane
at time $t=1.5$ with $\beta=1.0$.
The horizontal and vertical axes represent $x$ and $z$, respectively.
We draw the velocity of the fluid $\mbox{\boldmath $v$}(t,\mbox{\boldmath $x$})$
as an arrow (a vector) located at the point $\mbox{\boldmath $x$}$.
The lengths of arrows represent only just the ratios
of their norms compared with each other.}
\label{Figure04}
\end{figure}

Figures~\ref{Figure02}, \ref{Figure03} and \ref{Figure04} show the velocity
$\mbox{\boldmath $v$}(t,x,y,z)$ given by Eq.~(\ref{evolution-fluid-field-velocity-C})
on the $xz$-plane for $|\mbox{\boldmath $x$}|\leq 1$ and $z\leq 0$
with $\beta=1.0$
at $t=0.5$, $1.0$ and $1.5$, respectively.
Thus, we can think that Figs.~\ref{Figure02}, \ref{Figure03} and \ref{Figure04}
represent the tangent vectors of trajectories of $\mbox{\boldmath $S$}(t)$
at $t=0.5$, $1.0$ and $1.5$, respectively.
Looking at these figures,
we understand that the velocity of the fluid at any point at any time changes hard,
so that the particle of the fluid draws a very complicated trajectory.

In Sec.~\ref{section-physical-meanings},
we give physical meanings of $\rho$, $p$ and $\mbox{\boldmath $K$}$
defined in Eqs.~(\ref{equation-continuity}) and (\ref{Navier-Stokes-equations-0}).
In Appendix~\ref{section-appendix-A},
we consider how to build the Hamiltonian,
which generates the Navier-Stokes equations (\ref{Navier-Stokes-equations-0}).
Moreover, we consider the reason why the trajectory in Fig.~\ref{Figure01}
appears to intersect itself.
We explain that intersections never violate existence and uniqueness of solutions
for the Navier-Stokes equations.

\section{\label{section-physical-meanings}
Physical meanings of $\rho$, $p$ and $\mbox{\boldmath $K$}$}
In this section, we examine the physical meanings of
$\rho(t,\mbox{\boldmath $x$})$,
$p(t,\mbox{\boldmath $x$})$ and $\mbox{\boldmath $K$}(t,\mbox{\boldmath $x$})$,
which appear in Eqs.~(\ref{equation-continuity}) and (\ref{Navier-Stokes-equations-0}).
We introduce the compressible fluid
for explaining it to be equivalent to the evolution of the Bloch vector in
Eq.~(\ref{evolution-fluid-field}).
As a result of this procedure, we obtain these quantities,
as the density of the fluid,
the pressure and the external force per unit mass.
Because the compressible fluid itself is fictitious, we need to clarify their physical meanings.

First of all, using Eq.~(\ref{nabla-velocity-1}), we rewrite Eq.~(\ref{equation-continuity})
as follows:
\begin{equation}
\frac{\partial}{\partial t}\rho
+
\mbox{\boldmath $v$}\cdot\nabla\rho
+
\phi(t)\rho
=
0,
\label{equation-continuity-2-i}
\end{equation}
\begin{equation}
\phi(t)
=
\nabla\cdot\mbox{\boldmath $v$}
=
2\frac{\dot{L}_{1}(t)}{L_{1}(t)}+\frac{\dot{L}_{3}(t)}{L_{3}(t)}.
\label{equation-continuity-2-ii}
\end{equation}
To solve the partial differential equation (\ref{equation-continuity-2-i}),
we separate the density of the fluid $\rho(t,\mbox{\boldmath $x$})$
into two parts,
\begin{equation}
\rho(t,\mbox{\boldmath $x$})
=
R(t)W(\mbox{\boldmath $x$}).
\label{rho-separation-of-variables-1}
\end{equation}
Substitution of Eq.~(\ref{rho-separation-of-variables-1}) into Eq.~(\ref{equation-continuity-2-i})
yields
\begin{equation}
\frac{1}{R(t)}
\frac{\partial R(t)}{\partial t}
+
\phi(t)
+
\frac{1}{W(\mbox{\boldmath $x$})}
(\mbox{\boldmath $v$}\cdot\nabla)W(\mbox{\boldmath $x$})
=
0.
\label{equation-continuity-2-iii}
\end{equation}
Looking at Eq.~(\ref{equation-continuity-2-iii}), we notice the following facts.
On the one hand, the first and second terms depend on the variable $t$ only.
On the other hand, only the third term includes the variable $\mbox{\boldmath $x$}$.
However, because of Eq.~(\ref{evolution-fluid-field-velocity-C}),
the velocity is described as the function
$\mbox{\boldmath $v$}=\mbox{\boldmath $v$}(t,\mbox{\boldmath $x$})$,
so that the third term of Eq.~(\ref{equation-continuity-2-iii}) depends
on both $\mbox{\boldmath $x$}$ and $t$.
Putting these considerations together
and letting $\varphi(t)$ be an arbitrary function,
we can divide Eq.~(\ref{equation-continuity-2-iii})
into the following two differential equations:
\begin{equation}
\frac{1}{R(t)}\frac{\partial}{\partial t}R(t)+\phi(t)=\varphi(t),
\label{R-i}
\end{equation}
\begin{equation}
\frac{1}{W(\mbox{\boldmath $x$})}(\mbox{\boldmath $v$}\cdot\nabla)W(\mbox{\boldmath $x$})=-\varphi(t).
\label{W-i}
\end{equation}

Then, let us solve Eq.~(\ref{W-i}) first.
Dividing $W(\mbox{\boldmath $x$})$ into three parts as
\begin{equation}
W(\mbox{\boldmath $x$})=X(x)Y(y)Z(z),
\label{W-dividing-three-parts-XYZ}
\end{equation}
and taking care of Eq.~(\ref{evolution-fluid-field-velocity-C}),
we rewrite Eq.~(\ref{W-i}) as follows:
\begin{equation}
v_{1}(t,x)\frac{X'(x)}{X(x)}
+
v_{2}(t,y)\frac{Y'(y)}{Y(y)}
+
v_{3}(t,z)\frac{Z'(z)}{Z(z)}
=
-\varphi(t),
\label{differential-equation-XYZ-0}
\end{equation}
where
\begin{equation}
\mbox{\boldmath $v$}(t,\mbox{\boldmath $x$})
=
\left(
\begin{array}{l}
v_{1}(t,x) \\
v_{2}(t,y) \\
v_{3}(t,z)
\end{array}
\right),
\label{velocity-XYZ-1}
\end{equation}
and
\begin{eqnarray}
v_{1}(t,x)
&=&
\frac{\dot{L}_{1}(t)}{L_{1}(t)}x, \nonumber \\
v_{2}(t,y)
&=&
\frac{\dot{L}_{1}(t)}{L_{1}(t)}y, \nonumber \\
v_{3}(t,z)
&=&
\frac{\dot{L}_{3}(t)}{L_{3}(t)}[z-L_{4}(t)]+\dot{L}_{4}(t).
\label{velocity-elements-XYZ-2}
\end{eqnarray}
Moreover, we divide Eq.~(\ref{differential-equation-XYZ-0})
into the following three differential equations,
\begin{eqnarray}
v_{1}(t,x)\frac{X'(x)}{X(x)}
&=&
-\varphi_{1}(t),
\label{differential-equations-X-3} \\
v_{2}(t,y)\frac{Y'(y)}{Y(y)}
&=&
-\varphi_{2}(t),
\label{differential-equations-Y-3} \\
v_{3}(t,z)\frac{Z'(z)}{Z(z)}
&=&
-\varphi_{3}(t),
\label{differential-equations-Z-3}
\end{eqnarray}
where
\begin{equation}
\varphi_{1}(t)+\varphi_{2}(t)+\varphi_{3}(t)=\varphi(t).
\label{equation-varphi-4}
\end{equation}

Next, using Eq.~(\ref{velocity-elements-XYZ-2}),
we rewrite Eq.~(\ref{differential-equations-X-3}) as
\begin{equation}
x\frac{X'(x)}{X(x)}
=
-\frac{L_{1}(t)}{\dot{L}_{1}(t)}\varphi_{1}(t).
\label{differential-equation-X-5}
\end{equation}
The left-hand side of Eq.~(\ref{differential-equation-X-5}) depends only on $x$
and
the right-hand side of Eq.~(\ref{differential-equation-X-5}) depends only on $t$.
Thus, they have to be equal to an arbitrary constant $(-C_{1})$,
and we obtain
\begin{eqnarray}
xX'(x)=-C_{1}X(x), \label{differential-equation-X-6} \\
C_{1}\dot{L}_{1}(t)=L_{1}(t)\varphi_{1}(t).
\label{differential-equation-varphi-1-6}
\end{eqnarray}
From Eq.~(\ref{differential-equation-X-6}), we obtain
\begin{equation}
X(x)=D_{1}x^{-C_{1}},
\label{solution-X-7}
\end{equation}
where $D_{1}$ is an arbitrary constant.
In a similar way, from Eq.~(\ref{differential-equations-Y-3}), we obtain
\begin{eqnarray}
Y(y)=D_{2}y^{-C_{2}}, \nonumber \\
C_{2}\dot{L}_{1}(t)=L_{1}(t)\varphi_{2}(t),
\label{solution-Y-8}
\end{eqnarray}
where $C_{2}$ and $D_{2}$ are arbitrary constants.

Next, we think about Eq.~(\ref{differential-equations-Z-3}).
Using Eq.~(\ref{velocity-elements-XYZ-2}),
we rewrite Eq.~(\ref{differential-equations-Z-3}) as follows:
\begin{equation}
\frac{Z'(z)}{Z(z)}
=
-
\frac{\varphi_{3}(t)}{[\dot{L}_{3}(t)/L_{3}(t)][z-L_{4}(t)]+\dot{L}_{4}(t)}.
\label{differential-equation-Z-9}
\end{equation}
The left-hand side of Eq.~(\ref{differential-equation-Z-9}) depends only on $z$.
Thus, we have to let the right-hand side of Eq.~(\ref{differential-equation-Z-9})
not rely on the variable $t$,
which is independent of $z$.
This implies $\varphi_{3}(t)=0$.
Hence, we obtain $Z(z)=D_{3}$, where $D_{3}$ is an arbitrary constant.

Putting these results together, we obtain
\begin{eqnarray}
W(\mbox{\boldmath $x$})
&=&
D_{1}D_{2}D_{3}x^{-C_{1}}y^{-C_{2}}, \nonumber \\
\varphi(t)&=&(C_{1}+C_{2})\frac{\dot{L}_{1}(t)}{L_{1}(t)}.
\label{solutions-W-varphi-10}
\end{eqnarray}
Because of Eq.~(\ref{rho-separation-of-variables-1}),
we obtain $D_{1}D_{2}D_{3}\neq 0$,
so that we can avoid letting the density of the fluid
$\rho(t,\mbox{\boldmath $x$})[=R(t)W(\mbox{\boldmath $x$})]$
be equal to zero at every point $\mbox{\boldmath $x$}$ at every time $t$.
Moreover, we have to put $C_{1}\leq 0$ and $C_{2}\leq 0$
because we do not want to let $W(\mbox{\boldmath $x$})$ be singular at $x=0$ or $y=0$.
Similarly, because we assume that $W(\mbox{\boldmath $x$})$ converges to a finite value
in the limits of $|x|\to +\infty$ and $|y|\to +\infty$,
we put $C_{1}\geq 0$ and $C_{2}\geq 0$.
Hence, from the above discussions,
we set $C_{1}=C_{2}=0$.
Then, from Eq.~(\ref{solutions-W-varphi-10}),
we obtain $W(\mbox{\boldmath $x$})=W_{0}(=\mbox{constant})$
and $\varphi(t)=0$.
Therefore, from Eq.~(\ref{R-i}),
we arrive at
\begin{equation}
\frac{\partial}{\partial t}R(t)+\phi(t)R(t)=0.
\label{differential-equation-R-final-0}
\end{equation}

Taking care of $L_{1}(0)=L_{3}(0)=1$
for Eq.~(\ref{definition-L1-L3-L4})
and Eq.~(\ref{equation-continuity-2-ii}),
we obtain the solution of Eq.~(\ref{differential-equation-R-final-0}) as
\begin{eqnarray}
R(t)
&=&
R(0)\exp[-\int_{0}^{t}ds\phi(s)] \nonumber \\
&=&
\frac{R(0)}{L_{1}(t)^{2}L_{3}(t)}.
\end{eqnarray}
Therefore, we can write down $\rho(t,\mbox{\boldmath $x$})$ as follows:
\begin{equation}
\rho(t,\mbox{\boldmath $x$})=\rho(t)=\frac{\rho(0)}{L_{1}(t)^{2}L_{3}(t)}.
\label{rho-i}
\end{equation}
Equation~(\ref{rho-i}) tells us that the density of the fluid $\rho$ depends
only on $t$, and it does not rely on $\mbox{\boldmath $x$}$.

Next, we think around the pressure $p(t,\mbox{\boldmath $x$})$.
First, we have to carefully notice that we cannot determine $p(t,\mbox{\boldmath $x$})$
only by the Navier-Stokes equations.
To derive $p(t,\mbox{\boldmath $x$})$ in the fluid precisely,
we need an equation of state of the fluid given by classical thermodynamics.
To let the discussion be simple,
we assume the fluid to be barotropic, such as an ideal gas
\cite{Childress2009}.
For a barotropic fluid, the pressure is a function of the density alone as
\begin{equation}
p=f(\rho).
\label{barotropic-fluid-0}
\end{equation}
From Eqs.~(\ref{rho-i}) and (\ref{barotropic-fluid-0}), we obtain the pressure as
\begin{equation}
p(t,\mbox{\boldmath $x$})=f(\rho(t)).
\label{barotropic-fluid-1}
\end{equation}
Thus, we can conclude that the pressure $p(t,\mbox{\boldmath $x$})[=p(t)]$ depends only on $t$
and it can never be a function of $\mbox{\boldmath $x$}$.
Thus, we obtain
\begin{equation}
\nabla p=\mbox{\boldmath $0$}.
\label{barotropic-fluid-2}
\end{equation}
Putting these discussions and Eq.~(\ref{potential-fluid-field-D}) together,
we can rewrite Eq.~(\ref{Navier-Stokes-equations-0}) as follows:
\begin{equation}
\frac{\partial \mbox{\boldmath $v$}}{\partial t}
+
(\mbox{\boldmath $v$}\cdot\nabla)\mbox{\boldmath $v$}
=
\mbox{\boldmath $K$}.
\label{external-force-dash}
\end{equation}

From Eq.~(\ref{external-force-dash}), we can compute the external force per unit mass.
Substitution of Eq.~(\ref{evolution-fluid-field-velocity-C}) into Eq.~(\ref{external-force-dash})
yields an explicit form of $\mbox{\boldmath $K$}$ as
\begin{equation}
\mbox{\boldmath $K$}
=
\left(
\begin{array}{c}
{[}\ddot{L}_{1}(t)/L_{1}(t){]}x \\
{[}\ddot{L}_{1}(t)/L_{1}(t){]}y \\
{[}\ddot{L}_{3}(t)/L_{3}(t){]}[z-L_{4}(t)]+\ddot{L}_{4}(t)
\end{array}
\right).
\label{external-force-dash2}
\end{equation}
Equations~(\ref{external-force-dash}) and (\ref{external-force-dash2}) mean
that only the external force per unit mass $\mbox{\boldmath $K$}$ causes and affects the dynamics
of the velocity of the fluid $\mbox{\boldmath $v$}(t,\mbox{\boldmath $x$})$.

\section{\label{section-discussions}Discussions}
In this paper,
we rewrite the Bloch vector in the thermal JCM as the compressible inviscid fluid
with zero vorticity.
As shown in Sec.~\ref{section-equivalence-Bloch-vector-compressible-fluid},
the quasiperiodicity of the Bloch vector causes nontrivial dynamics of the compressible fluid.
Such an explicit connection between a quantum mechanical model and a classical field theory
surprises us.

Here, let us consider realistic fluid that displays the dynamical evolution of the Bloch vector.
From the results of Sec.~\ref{section-equivalence-Bloch-vector-compressible-fluid},
the fluid has to be compressible and inviscid.
Moreover, it has no vorticity.
Thus, we imagine that a sphere container is filled with a gas.
Because the norm of the Bloch vector is equal to or less than unity,
the radius of the sphere is equal to unity.
We assume the velocity of the flow of the gas in the sphere is very high,
so that the gas becomes significantly compressible.
The movements of the flow are caused by the external force per unit mass $\mbox{\boldmath $K$}$
given by Eq.~(\ref{external-force-dash2}).
In Eq.~(\ref{fictitious-fluid-0}) of Appendix~\ref{section-appendix-A},
we show
$\mbox{\boldmath $K$}=-\nabla K$ holds.
This implies that $\mbox{\boldmath $K$}$ is a conservative body force.
Because the gas is barotropic ideal fluid under the conservative body force,
we can apply Kelvin's theorem (the law of conservation of circulation) to it \cite{Landau1987,Lamb1932,Childress2009}.
(Because the gas is inviscid, we can regard it as the ideal fluid.)
Thus, if the fluid in the container has no vorticity at initial time,
it never has any vorticity at arbitrary time.

During the time evolution,
both the direction and the norm of the Bloch vector change hard at random.
We draw an analogy between the compressible fluid and the Bloch vector under the thermal JCM.
We can regard the compressible flow as the origin of this randomness.
Then,
the following question arises.
Can we rewrite other quantum mechanical models as fluid dynamic models?
This question remains to be answered.

As mentioned in Sec.~\ref{section-introduction},
we obtain the Bloch vector by tracing out the Hilbert space of the cavity field.
Because entanglement between the atom and the cavity field is removed,
the dynamics of the Bloch vector is described with a hidden-variable model governed by classical deterministic logic.
However, the Bloch vector still contains quantum nature, for example,
the discreteness of states.
That is to say, discreteness of the number of photons remains as the summation $\sum_{n}$ in $L_{1}(t)$, $L_{3}(t)$ and $L_{4}(t)$
given by Eq.~(\ref{definition-L1-L3-L4}).
This summation realizes a superposition of terms that are characterized with incommensurate angular frequencies.
Hence, thermal disorder induced by the cavity field in thermal equilibrium drives the Bloch vector to draw its complex trajectory.
As a result of these considerations, we understand that the dynamics of the thermal Bloch vector is generated
by effects of both quantum discreteness and thermal disorder.

Since the JCM appeared,
it has been providing a lot of information concerning quantum nature to us.
However, the authors think the JCM reveals us more and more amazing properties
in the near future.

\appendix
\section{\label{section-appendix-A}
The Hamiltonian mechanics for the compressible fluid}
In this section,
we consider how to build the Hamiltonian,
which yields dynamics of the compressible fluid being equivalent to the Bloch vector.
In Sec.~\ref{section-equivalence-Bloch-vector-compressible-fluid},
we show that the dynamics of the Bloch vector can be described with the Navier-Stokes equations
for the inviscid compressible fluid with zero vorticity.
In the following, we construct the Hamiltonian that provides the Navier-Stokes equations.
Moreover, we consider the reason why the trajectory in Fig.~\ref{Figure01} appears to intersect itself.
At the end of this section,
we explain that the trajectory never intersects itself in the phase space.
Thus, the intersections in Fig.~\ref{Figure01}
are the result of a projection operation
and they never violate existence and uniqueness of solutions for the Navier-Stokes equations.

Now, let us consider how to construct the Hamiltonian for a fluid particle
\cite{Salmon1988}.
First, we prepare a three-dimensional real vector
$\mbox{\boldmath $a$}=(a_{1},a_{2},a_{3})$ as a particle label at time $\tau$.
The vector $\mbox{\boldmath $a$}$ is called curvilinear labelling co-ordinates.
They are co-ordinates fixed on the fluid particle.
The time variable $\tau$ is a proper time,
which is shown by a clock attached to the fluid particle.
Thus, the fluid particle is determined completely by $(\tau,\mbox{\boldmath $a$})$.

Let $\mbox{\boldmath $x$}(\tau,\mbox{\boldmath $a$})$ be the location of the fluid particle
identified by $\mbox{\boldmath $a$}$ and $\tau$.
Then, the vector $\mbox{\boldmath $a$}$ specifies the mass element of the fluid particle,
so that the Jacobian determinant of $\mbox{\boldmath $a$}$ gives a mass-density of the fluid.
The labelling co-ordinates $\mbox{\boldmath $a$}$ remain constant following the motion of the fluid particle.

Here, we describe a mass element of the fluid as
\begin{equation}
d(\mbox{mass})=da_{1}da_{2}da_{3}.
\end{equation}
The partial differentiation of the proper time $\tau$ is equal to the material derivative as
\begin{equation}
\tau = t,
\quad
\frac{\partial}{\partial\tau}
=
\frac{D}{Dt}
=
\frac{\partial}{\partial t}+\mbox{\boldmath $v$}\cdot\nabla,
\label{material-derivative-0}
\end{equation}
where
\begin{equation}
\mbox{\boldmath $v$}
=
(u,v,w)
=
(\frac{\partial x}{\partial\tau},\frac{\partial y}{\partial\tau},\frac{\partial z}{\partial\tau}).
\label{material-derivative-1}
\end{equation}
Then, the mass-density of the fluid is given by
\begin{equation}
\rho
=
\frac{\partial(a_{1},a_{2},a_{3})}{\partial(x,y,z)}
=
\frac{\partial(\mbox{\boldmath $a$})}{\partial(\mbox{\boldmath $x$})}.
\end{equation}
Moreover,
we define the specific volume of the fluid as
\begin{equation}
\alpha
=
\frac{1}{\rho}
=
\frac{\partial(\mbox{\boldmath $x$})}{\partial(\mbox{\boldmath $a$})}.
\label{specific-volume-0}
\end{equation}

From the above notations,
we obtain
\begin{eqnarray}
\frac{\partial\alpha}{\partial\tau}
&=&
\frac{\partial(u,y,z)}{\partial(a_{1},a_{2},a_{3})}
+
\frac{\partial(x,v,z)}{\partial(a_{1},a_{2},a_{3})}
+
\frac{\partial(x,y,w)}{\partial(a_{1},a_{2},a_{3})} \nonumber \\
&=&
\frac{\partial(\mbox{\boldmath $x$})}{\partial(\mbox{\boldmath $a$})}
[
\frac{\partial(u,y,z)}{\partial(x,y,z)}
+
\frac{\partial(x,v,z)}{\partial(x,y,z)}
+
\frac{\partial(x,y,w)}{\partial(x,y,z)}
] \nonumber \\
&=&
\alpha
(\frac{\partial u}{\partial x}+\frac{\partial v}{\partial y}+\frac{\partial w}{\partial z}) \nonumber \\
&=&
\alpha\nabla\cdot\mbox{\boldmath $v$},
\end{eqnarray}
and
\begin{equation}
\frac{\partial}{\partial\tau}
\frac{1}{\rho}
=
-
\frac{1}{\rho^{2}}
\frac{\partial\rho}{\partial\tau}
=
\frac{1}{\rho}\nabla\cdot\mbox{\boldmath $v$},
\end{equation}
so that we arrive at the continuity equation,
\begin{equation}
\frac{\partial\rho}{\partial\tau}+\rho\nabla\cdot\mbox{\boldmath $v$}=0.
\label{continuity-equation-0}
\end{equation}
Using Eq.~(\ref{material-derivative-0}),
we can rewrite Eq.~(\ref{continuity-equation-0}) as
\begin{eqnarray}
&&
\frac{\partial \rho}{\partial t}
+
(\mbox{\boldmath $v$}\cdot\nabla)\rho
+
\rho\nabla\cdot\mbox{\boldmath $v$} \nonumber \\
&=&
\frac{\partial\rho}{\partial t}
+
\nabla\cdot(\rho\mbox{\boldmath $v$}) \nonumber \\
&=&
0.
\end{eqnarray}

Next, we define the Lagrangian as follows:
\begin{equation}
L
=
\int \!\!\! \int \!\!\! \int
d^{3}\mbox{\boldmath $a$}
[
\frac{1}{2}(\frac{\partial\mbox{\boldmath $x$}}{\partial\tau})^{2}
-
E(\frac{1}{\rho})
-
K(\mbox{\boldmath $x$},t)
],
\label{Lagrangian-fluid-particle-0}
\end{equation}
where $E(1/\rho)$ and $K(t,\mbox{\boldmath $x$})$ represent the specific internal energy
and the potential for an external force, respectively.
We describe Hamilton's principle as
\begin{equation}
\delta\int L d\tau = 0,
\end{equation}
where $\delta$ represents an arbitrary independent variation
$\delta\mbox{\boldmath $x$}(\tau,\mbox{\boldmath $a$})$
for the fluid particle locations.
Then, we can calculate $\delta L$ as follows:
\begin{eqnarray}
\delta L
&=&
\delta
\int d\tau \int \!\!\! \int \!\!\! \int d^{3}\mbox{\boldmath $a$}
[
\frac{1}{2}
\frac{\partial \mbox{\boldmath $x$}}{\partial\tau}\cdot\frac{\partial \mbox{\boldmath $x$}}{\partial\tau}
-
E(\frac{1}{\rho})
-
K(t,\mbox{\boldmath $x$})
] \nonumber \\
&=&
\int d\tau \int \!\!\! \int \!\!\! \int d^{3}\mbox{\boldmath $a$}
[
-
\frac{\partial^{2} \mbox{\boldmath $x$}}{\partial\tau^{2}}\cdot\delta\mbox{\boldmath $x$}
-
\frac{\partial E(\alpha)}{\partial\alpha}
\delta\frac{\partial(\mbox{\boldmath $x$})}{\partial(\mbox{\boldmath $a$})}
-
\frac{\partial K}{\partial\mbox{\boldmath $x$}}\cdot\delta\mbox{\boldmath $x$}
].
\label{delta-Lagrangian-0}
\end{eqnarray}
To derive the above equation, we use Eq.~(\ref{specific-volume-0}).

Here, we introduce a convenient formula.
For an arbitrary function $F=F(t,\mbox{\boldmath $x$})=F(\tau,\mbox{\boldmath $a$})$,
the following relation holds:
\begin{eqnarray}
&&
\int \!\!\! \int \!\!\! \int d^{3}\mbox{\boldmath $a$}
F\delta\frac{\partial(\mbox{\boldmath $x$})}{\partial(\mbox{\boldmath $a$})} \nonumber \\
&=&
\int \!\!\! \int \!\!\! \int d^{3}\mbox{\boldmath $a$}
F
[
\frac{\partial(\delta x,y,z)}{\partial(a_{1},a_{2},a_{3})}
+
\frac{\partial(x,\delta y,z)}{\partial(a_{1},a_{2},a_{3})}
+
\frac{\partial(x,y,\delta z)}{\partial(a_{1},a_{2},a_{3})}
] \nonumber \\
&=&
\int \!\!\! \int \!\!\! \int d^{3}\mbox{\boldmath $a$}
F
\frac{\partial(\mbox{\boldmath $x$})}{\partial(\mbox{\boldmath $a$})}
[
\frac{\partial(\delta x,y,z)}{\partial(x,y,z)}
+
\frac{\partial(x,\delta y,z)}{\partial(x,y,z)}
+
\frac{\partial(x,y,\delta z)}{\partial(x,y,z)}
] \nonumber \\
&=&
\int \!\!\! \int \!\!\! \int d^{3}\mbox{\boldmath $a$}
F
\frac{\partial(\mbox{\boldmath $x$})}{\partial(\mbox{\boldmath $a$})}
(\nabla\cdot\delta \mbox{\boldmath $x$}) \nonumber \\
&=&
\int \!\!\! \int \!\!\! \int d^{3}\mbox{\boldmath $x$}
F
(\nabla\cdot\delta \mbox{\boldmath $x$}) \nonumber \\
&=&
-
\int \!\!\! \int \!\!\! \int d^{3}\mbox{\boldmath $x$}
\delta\mbox{\boldmath $x$}\cdot\nabla F
+
\int \!\!\! \int_{\partial V}
F\delta\mbox{\boldmath $x$}\cdot d\mbox{\boldmath $\sigma$} \nonumber \\
&=&
-
\int \!\!\! \int \!\!\! \int d^{3}\mbox{\boldmath $x$}
\delta\mbox{\boldmath $x$}\cdot\nabla F.
\end{eqnarray}
In the above derivation,
we drop the second term as a surface integral on the two-dimensional boundary.

Thus, we can rewrite Eq.~(\ref{delta-Lagrangian-0}) as follows:
\begin{equation}
\delta L
=
-
\int d\tau
\int \!\!\! \int \!\!\! \int d^{3}\mbox{\boldmath $x$}
\rho
(
\frac{\partial^{2}\mbox{\boldmath $x$}}{\partial \tau^{2}}
+
\frac{1}{\rho}\nabla p
+
\nabla K
)
\cdot
\delta\mbox{\boldmath $x$},
\end{equation}
where we use the thermodynamical relation,
\begin{equation}
p
=
-
\frac{\partial E(\alpha)}{\partial\alpha},
\end{equation}
and $d^{3}\mbox{\boldmath $a$}=d^{3}\mbox{\boldmath $x$}\rho$.

Hence, we obtain the following equations of motion:
\begin{equation}
\frac{\partial^{2}\mbox{\boldmath $x$}}{\partial\tau^{2}}
+
\frac{1}{\rho}\nabla p
+
\nabla K
=
0.
\end{equation}
Then, using Eqs.~(\ref{material-derivative-0}) and (\ref{material-derivative-1}),
we can rewrite the above equations as
\begin{equation}
\frac{D\mbox{\boldmath $v$}}{Dt}
+
\frac{1}{\rho}
\nabla p
+
\nabla K
=
0.
\end{equation}
Consequently, we obtain the following equations of motion:
\begin{eqnarray}
\frac{\partial \mbox{\boldmath $v$}}{\partial t}
+
(\mbox{\boldmath $v$}\cdot\nabla)\mbox{\boldmath $v$}
+
\frac{1}{\rho}\nabla p
+
\nabla K
&=&
0, \label{equations-motion-1} \\
\frac{\partial \rho}{\partial t}
+
\nabla\cdot(\rho\mbox{\boldmath $v$})
&=&
0, \label{equations-motion-2} \\
\frac{\partial E(1/\rho)}{\partial(1/\rho)}
+
p
&=&
0. \label{equations-motion-3}
\end{eqnarray}

Comparing Eq.~(\ref{equations-motion-1}) with Eq.~(\ref{Navier-Stokes-equations-0}),
we obtain
\begin{eqnarray}
\nabla K
&=&
-
\mbox{\boldmath $K$}, \nonumber \\
\mbox{\boldmath $v$}
&=&
\nabla \Phi, \nonumber \\
K
&=&
-
\frac{\ddot{L}_{1}(t)}{L_{1}(t)}
(\frac{x^{2}}{2}+\frac{y^{2}}{2})
-
\frac{\ddot{L}_{3}(t)}{L_{3}(t)}
[
\frac{z^{2}}{2}-zL_{4}(t)
]
-
\ddot{L}_{4}(t)z,
\label{fictitious-fluid-0}
\end{eqnarray}
and
$\rho$, $\mbox{\boldmath $K$}$, $p$ and $\Phi$ are given
by Eqs.~(\ref{rho-i}), (\ref{external-force-dash2}), (\ref{barotropic-fluid-1})
and (\ref{potential-fluid-field-B}), respectively.

Because the Lagrangian density given by Eq.~(\ref{Lagrangian-fluid-particle-0})
is expressed as a function of $(t,\mbox{\boldmath $x$})$,
some might consider the number of degrees of freedom to be equal to three.
However, the integration over the labelling co-ordinates $\mbox{\boldmath $a$}$,
that is,
$\int \!\! \int \!\! \int d^{3}\mbox{\boldmath $a$}$ in Eq.~(\ref{Lagrangian-fluid-particle-0}),
gives an infinite number of degrees of freedom to the system.
The Lagrangian in Eq.~(\ref{Lagrangian-fluid-particle-0}) is a sum of Lagrangians,
each of which determines dynamics of each single fluid particle.
Here, let us rewrite the Lagrangian with a field of the flow $\varphi (t,\mbox{\boldmath $x$})$
and a field of the density $\rho (t,\mbox{\boldmath $x$})$ as
\begin{equation}
L
=
\int \!\!\! \int \!\!\! \int d^{3}\mbox{\boldmath $x$}
[
\frac{1}{2}
\rho(\nabla\varphi)\cdot(\nabla\varphi)
-
\tilde{E}(\rho)
-
K(t,\mbox{\boldmath $x$})\rho
],
\end{equation}
where
\begin{equation}
\frac{\partial^{2}\tilde{E}}{\partial\rho^{2}}
=
\frac{1}{\rho}
\frac{\partial p(\rho)}{\partial \rho}.
\end{equation}
In the derivation of the above Lagrangian,
to replace labelling co-ordinates $\mbox{\boldmath $a$}(t,\mbox{\boldmath $x$})$
with the field of the density $\rho(t,\mbox{\boldmath $x$})$
as $d^{3}\mbox{\boldmath $a$}=d^{3}\mbox{\boldmath $x$}\rho$ is essential.

Then, the Hamiltonian is given by
\begin{equation}
H
=
\int \!\!\! \int \!\!\! \int d^{3}\mbox{\boldmath $x$}
[
\frac{1}{2}
\rho(\nabla\varphi)\cdot(\nabla\varphi)
+
\tilde{E}(\rho)
+
K(t,\mbox{\boldmath $x$})\rho
].
\label{Hamiltonian-fluid-mechanics-0}
\end{equation}
Hence, we can derive equations of motion as
\begin{equation}
\frac{\partial\rho}{\partial t}
=
\frac{\delta H}{\delta\varphi}
=
-
\nabla\cdot(\rho\nabla\varphi),
\label{equation-motion-fluid-field-1}
\end{equation}
\begin{equation}
\frac{\partial\varphi}{\partial t}
=
-
\frac{\delta H}{\delta\rho}
=
-
\frac{1}{2}(\nabla\varphi)\cdot(\nabla\varphi)
-
\tilde{E}'(\rho)
-
K(t,\mbox{\boldmath $x$}),
\label{equation-motion-fluid-field-2}
\end{equation}
\begin{equation}
\mbox{\boldmath $v$}
=
\nabla\varphi.
\label{equation-motion-fluid-field-3}
\end{equation}
From Eqs.~(\ref{equation-motion-fluid-field-1}) and (\ref{equation-motion-fluid-field-3}),
we obtain the continuity equation (\ref{equations-motion-2}).
Moreover, from Eqs.~(\ref{equation-motion-fluid-field-2}) and (\ref{equation-motion-fluid-field-3}),
we obtain
\begin{equation}
\frac{\partial \mbox{\boldmath $v$}}{\partial t}
=
-
(\mbox{\boldmath $v$}\cdot\nabla)\mbox{\boldmath $v$}
-
\tilde{E}''(\rho)\nabla\rho
-
\nabla K.
\end{equation}
Because
\begin{equation}
\tilde{E}''(\rho)\nabla\rho
=
\frac{1}{\rho}
\frac{\partial p(\rho)}{\partial\rho}\nabla\rho
=
\frac{1}{\rho}\nabla p(\rho),
\end{equation}
we can derive the Navier-Stokes equations,
\begin{equation}
\frac{\partial\mbox{\boldmath $v$}}{\partial t}
=
-
(\mbox{\boldmath $v$}\cdot\nabla)\mbox{\boldmath $v$}
-
\frac{1}{\rho}\nabla p
+
\mbox{\boldmath $K$}.
\end{equation}

\begin{figure}
\begin{center}
\includegraphics[scale=1.0]{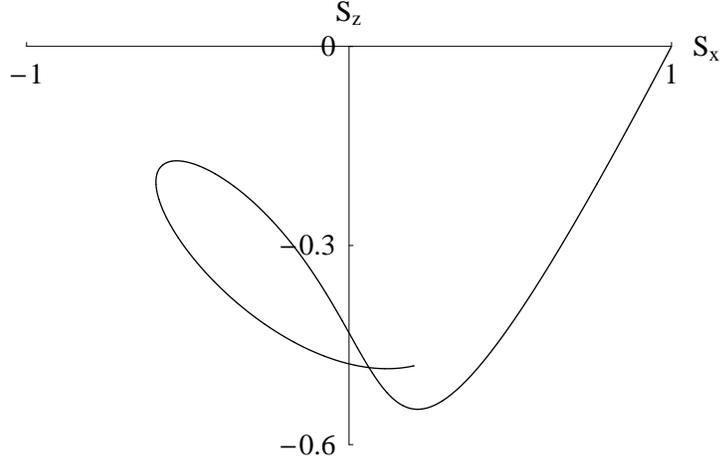}
\end{center}
\caption{The trajectory of the Bloch vector $\mbox{\boldmath $S$}(t)$
given
by Eqs.~(\ref{definition-L1-L3-L4}) and (\ref{Bloch-xz-plane})
for
$0\leq t\leq 5$,
where $\mbox{\boldmath $S$}(0)=(1,0,0)$ and $\beta=1.0$.
The horizontal and vertical axes represent $S_{x}$ and $S_{z}$, respectively.
The trajectory intersects itself at $t_{1}=1.644$ and $t_{2}=4.809$.}
\label{Figure05}
\end{figure}

In Fig.~\ref{Figure05},
the trajectory of the Bloch vector intersects itself.
As explained in Ref.~\cite{Azuma2014},
for the quasiperiodic system,
the trajectory of the motion never intersects itself
during a finite time interval in the phase space
because of incommensurate angular frequencies.
The quasiperiodic trajectory becomes dense on the surface (submanifold)
specified by the integrals of motion in the phase space.
This fact seems to contradict the intersection of the trajectory of the Bloch vector
in Fig.~\ref{Figure05}.
We examine this point with numerical calculations.

First, we think about a certain fluid particle labelled with $(\tau,\mbox{\boldmath $a$})$.
From the Lagrangian given by Eq.~(\ref{Lagrangian-fluid-particle-0}),
we understand that its generalized positions and momenta correspond to $\mbox{\boldmath $x$}(=\mbox{\boldmath $S$})$
and $\mbox{\boldmath $v$}(=\partial \mbox{\boldmath $x$}/\partial \tau)$,
respectively.
Thus, the phase space of the fluid particle $(\tau,\mbox{\boldmath $a$})$ is constructed with variables,
$\mbox{\boldmath $x$}(t)[=\mbox{\boldmath $S$}(t)]$
and
$\mbox{\boldmath $v$}(t,\mbox{\boldmath $x$})$.

Figure~\ref{Figure05} shows the trajectory of the Bloch vector
$\mbox{\boldmath $S$}(t)$
given
by
Eqs.~(\ref{definition-L1-L3-L4}) and (\ref{Bloch-xz-plane})
for
$0\leq t\leq 5$,
where $\mbox{\boldmath $S$}(0)=(1,0,0)$ and $\beta=1.0$.
Looking at Fig.~\ref{Figure05},
we notice that the trajectory intersects itself on
$(S_{x},S_{z})=(0.063\mbox{ }72,-0.4840)$
at
$t_{1}=1.644$ and $t_{2}=4.809$.
However, in the phase space,
the orbit of the motion never closes on itself at time variables $t_{1}$ and $t_{2}$.
We can confirm this fact as follows.
Regarding the system as the fluid particle being inviscid and compressible with zero vorticity,
we can understand dynamics of the system through the Lagrangian
given by Eq.~(\ref{Lagrangian-fluid-particle-0}).
Using Eq.~(\ref{evolution-fluid-field-velocity-C}),
the velocity vectors $(v_{x},v_{z})$ at $t_{1}$ and $t_{2}$ are given numerically as
\begin{eqnarray}
v_{x}(t_{1})=-0.3632,
&\quad\quad&
v_{z}(t_{1})=0.2681, \nonumber\\
v_{x}(t_{2})=-0.7494,
&\quad\quad&
v_{z}(t_{2})=-0.043\mbox{ }48.
\end{eqnarray}
Hence, in the phase space,
states of the system at $t_{1}$ and $t_{2}$ are clearly different from each other.

\end{document}